\documentclass[twocolumn,showpacs,preprintnumbers,amsmath,amssymb,prl,aps,superscriptaddress]{revtex4}
\usepackage{graphicx}
\usepackage{amsmath}
\usepackage{verbatim}

\newcommand{\fge}{f_{\mathrm{01}}}

\newcommand{\Ipi}[1]{I_{p,#1}}

\newcommand{\fc}{f_{\mathrm{c}}}

\newcommand{\Ec}{E_\mathrm{C}}
\newcommand{\Eth}{E_{\mathrm{Th}}}
\newcommand{\Ej}{E_{\mathrm{J}}}

\newcommand{\fbare}{f_{\mathrm{bare}}}

\newcommand{\fLO}{f_{\mathrm{LO}}}
\newcommand{\Vg}{V_{\mathrm{g}}}
\newcommand{\Tesla}{\mathrm{T}}

\newcommand{\cpr}{CPR}
\newcommand{\MHz}{\mathrm{MHz}}
\newcommand{\GHz}{\mathrm{GHz}}
\newcommand{\V}{\mathrm{V}}
\newcommand{\cm}{\mathrm{cm}}
\newcommand{\nm}{\mathrm{nm}}
\newcommand{\um}{\mu\mathrm{m}}
\newcommand{\nA}{\mathrm{nA}}
\newcommand{\us}{\mu\mathrm{s}}
\newcommand{\s}{\mathrm{s}}

\begin{document}
\title{Realization of microwave quantum circuits using hybrid superconducting-semiconducting nanowire Josephson elements}

\author{G.~de Lange}
\affiliation{QuTech and Kavli Institute of Nanoscience, Delft University of Technology, 2600 GA Delft, The Netherlands.}

\author{B. van Heck}
\affiliation{Instituut-Lorentz, Leiden University, 2300 RA Leiden, The Netherlands.}

\author{A. Bruno}

\author{D.~J. van Woerkom}

\author{A. Geresdi}
\affiliation{QuTech and Kavli Institute of Nanoscience, Delft University of Technology, 2600 GA Delft, The Netherlands.}

\author{S.~R.~Plissard}
\affiliation{Department of Applied Physics, Eindhoven University of Technology, 5600 MB Eindhoven, The Netherlands.}

\author{E.~P.~A.~M.~Bakkers}

\affiliation{QuTech and Kavli Institute of Nanoscience, Delft University of Technology, 2600 GA Delft, The Netherlands.}
\affiliation{Department of Applied Physics, Eindhoven University of Technology, 5600 MB Eindhoven, The Netherlands.}

\author{A.~R.~Akhmerov}

\author{L.~DiCarlo}
\affiliation{QuTech and Kavli Institute of Nanoscience, Delft University of Technology, 2600 GA Delft, The Netherlands.}

\date{\today}

\begin{abstract}
We report the realization of quantum microwave circuits using hybrid superconductor-semiconductor Josephson elements comprised of InAs nanowires contacted by NbTiN. Capacitively-shunted single elements behave as transmon qubits with electrically tunable transition frequencies. Two-element circuits also exhibit transmon-like behavior near zero applied flux, but behave as flux qubits at half the flux quantum,  where non-sinusoidal current-phase relations in the elements produce a double-well Josephson potential.  These hybrid Josephson elements are promising for applications requiring microwave superconducting circuits operating in magnetic field.
\end{abstract}


\maketitle

In superconducting circuits, macroscopic degrees of freedom like currents and voltages can exhibit quantum mechanical behavior. These circuits can behave as artificial atoms with discrete, anharmonic levels whose transitions can be driven coherently~\cite{Clarke08}. In the field of circuit quantum electrodynamics (cQED), these artificial atoms are coupled to resonators to perform microwave quantum optics in the solid state~\cite{Blais04, Wallraff04}. Over the last decade, cQED has also grown into a promising platform for quantum information processing, wherein the ground and first-excited levels of each atom serve as an effective qubit~\cite{Devoret13}. To date, implementations of superconducting quantum circuits have relied almost exclusively on aluminum/aluminum-oxide/aluminum (Al/AlOx/Al) tunnel junctions as the source of non-linearity without dissipation. However, many exciting applications require magnetic fields ($\sim 0.5~\Tesla$) at which superconductivity in aluminum is destroyed, calling for an alternative approach to realizing microwave artificial atoms.

Recent advances in materials development and nanowire (NW) growth have enabled the development of superconductor-semiconductor (super-semi) structures supporting coherent charge transport without dissipation~\cite{Doh05}, and providing signatures of Majorana bound states~\cite{Mourik12}. To date, super-semi-super Josephson elements (JEs) have been studied exclusively in quasi-DC transport~\cite{Jespersen09, Ziino13, Abay14, Sochnikov14}. Building microwave circuits operating in the quantum regime, in which transition energies between levels exceed the thermal energy, offers news ways to investigate the physics of hybrid super-semi structures using spectroscopy~\cite{Pekker13, Muller13,Ginossar14}.

\begin{figure}[h]
\centering
\includegraphics[width=\columnwidth]{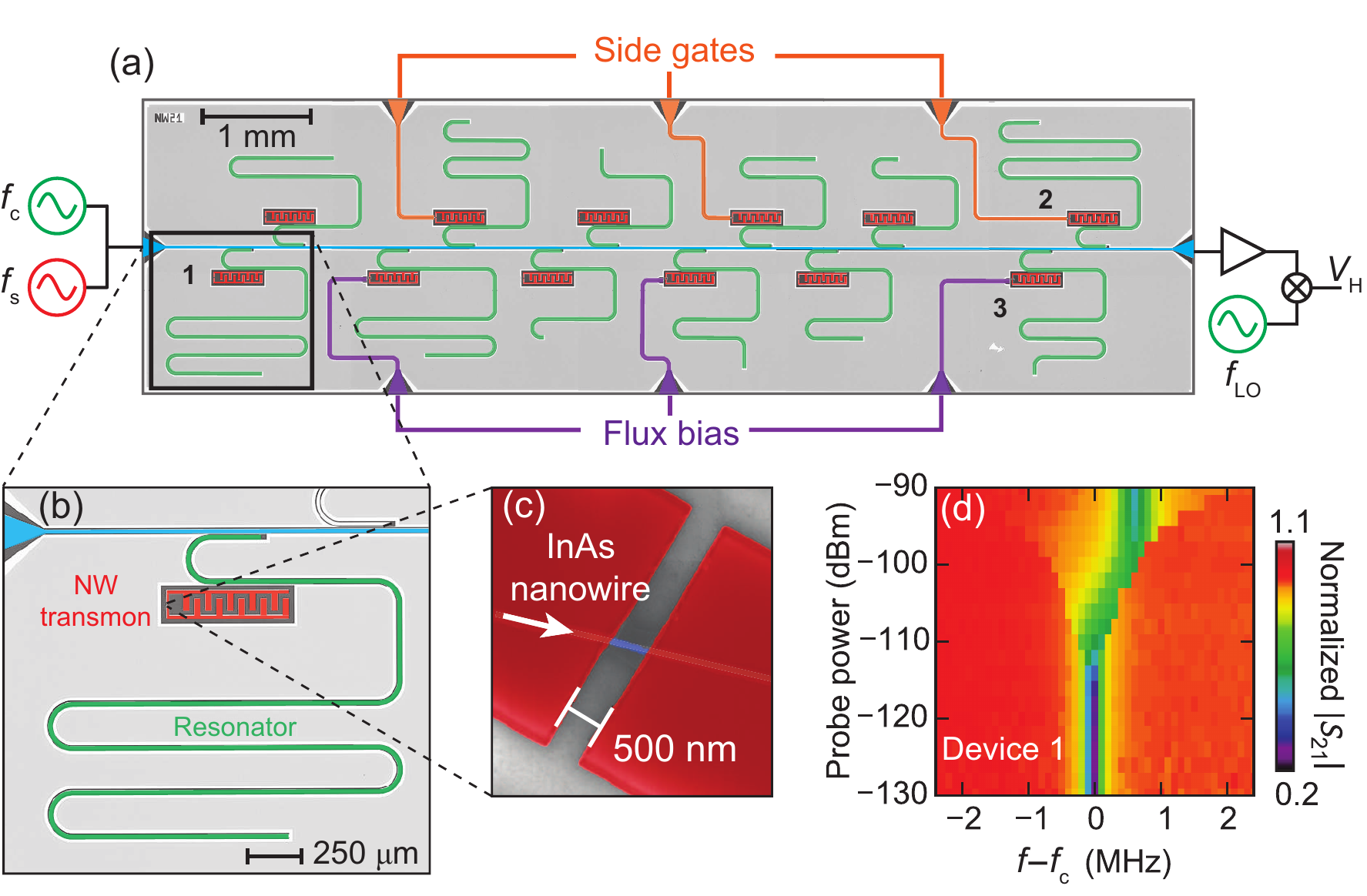}
\caption{Hybrid Josephson elements in cQED. (a) Overview of cQED chip allowing control and readout of NW circuits using dedicated resonators (green) coupled to a common feedline (blue). For readout, a microwave tone with frequency $\fc$ is applied near the fundamental of the resonator coupling to the NW circuit under study. The signal is amplified and downconverted to $1~\MHz$ using a local oscillator at $\fLO$ for subsequent digitization and processing. Additional controls on a subset of devices include side gates (orange) for electrostatic tuning of carrier density in the NW of single-junction devices, and short-circuited transmission lines (purple) for threading flux through the loops of split-junction devices. Measurements on devices 1-3 are described in the main text. See table S1 for an overview of the other devices. (b) Optical zoom-in of Device 1, containing a single-junction NW circuit (red). (c) Scanning electron microscope (SEM) image of an InAs NW (blue) contacted by NbTiN electrodes (red) separated by $500~\nm$. (d) Normalized feedline transmission as a function of readout power. The resonator shifts from $\fc = 3.9464~\GHz$ at single-photon level to $\fbare = 3.9470~\GHz$ above $ 10^5$ photons. This shift confirms the coupling of the resonator to a non-linear circuit.
}
\end{figure}

In this Letter, we report the realization of microwave-frequency cQED circuits made from hybrid JEs based on InAs NWs contacted by NbTiN. Capacitively shunted single JEs  behave as weakly anharmonic oscillators, or transmons~\cite{Koch07}, with transition frequencies tuneable by the field effect, i.e., voltage on a proximal side gate.  Double-element devices show similar transmon-like behavior at zero applied flux, but behave as flux qubits~\cite{Mooij99} near full frustration owing to a double-well Josephson potential arising from non-sinusoidal current-phase relations (\cpr s). We observe microwave driven transitions between states with oppositely flowing persistent currents, manifesting macroscopic quantum coherence. Their fabrication from magnetic-field compatible materials makes these JEs promising for applications requiring quantum microwave circuits withstanding magnetic fields up to  $~0.5~\Tesla$, such as the braiding of Majorana particles and the interfacing with coherent quantum memories based on polarized electron-spin ensembles.

Our chip [Fig.~1(a)] contains multiple capacitively shunted single and double NW JEs coupled to dedicated transmission-line resonators for control and readout using a common feedline [Fig.~1(b)]. The chip contains side gates for electrostatic tuning of some single-junction devices and current-bias lines for threading flux through the loops of split-junction devices.
We created each JE by deterministically placing an InAs NW between the leads of a pre-patterned NbTiN interdigitated capacitor (IDC) and contacting the NW to each lead in a subsequent NbTiN deposition. The charging energy $\Ec \approx h \times 300~\MHz$ of the devices is chosen much smaller than the estimated Josephson coupling energy $\Ej$ of the NW junction, as in conventional transmon devices~\cite{Koch07}, leading to a weakly anharmonic energy spectrum (energies $E_i$) of circuit plasma modes. We first verify the presence of the non-linear NW circuit by measuring the feedline transmission near the fundamental frequency of the coupled resonator [Fig.~1(d)].
The Jaynes-Cummings interaction leads to different resonator frequencies $\fc$ and $\fbare$ at single- and many-photon probe levels, respectively~\cite{Reed10, Bishop10}. We then search for the qubit transition frequency  $\fge$ of the NW circuit by monitoring feedline transmission at $\fc$ while sweeping a second tone~\cite{Schreier08} near the estimated frequency $\fbare + \left(\fbare -\fc \right)/g^{2}$, where $g$ is the coupling strength between NW circuit and resonator.

\begin{figure}[h]
\centering
\includegraphics[width=\columnwidth]{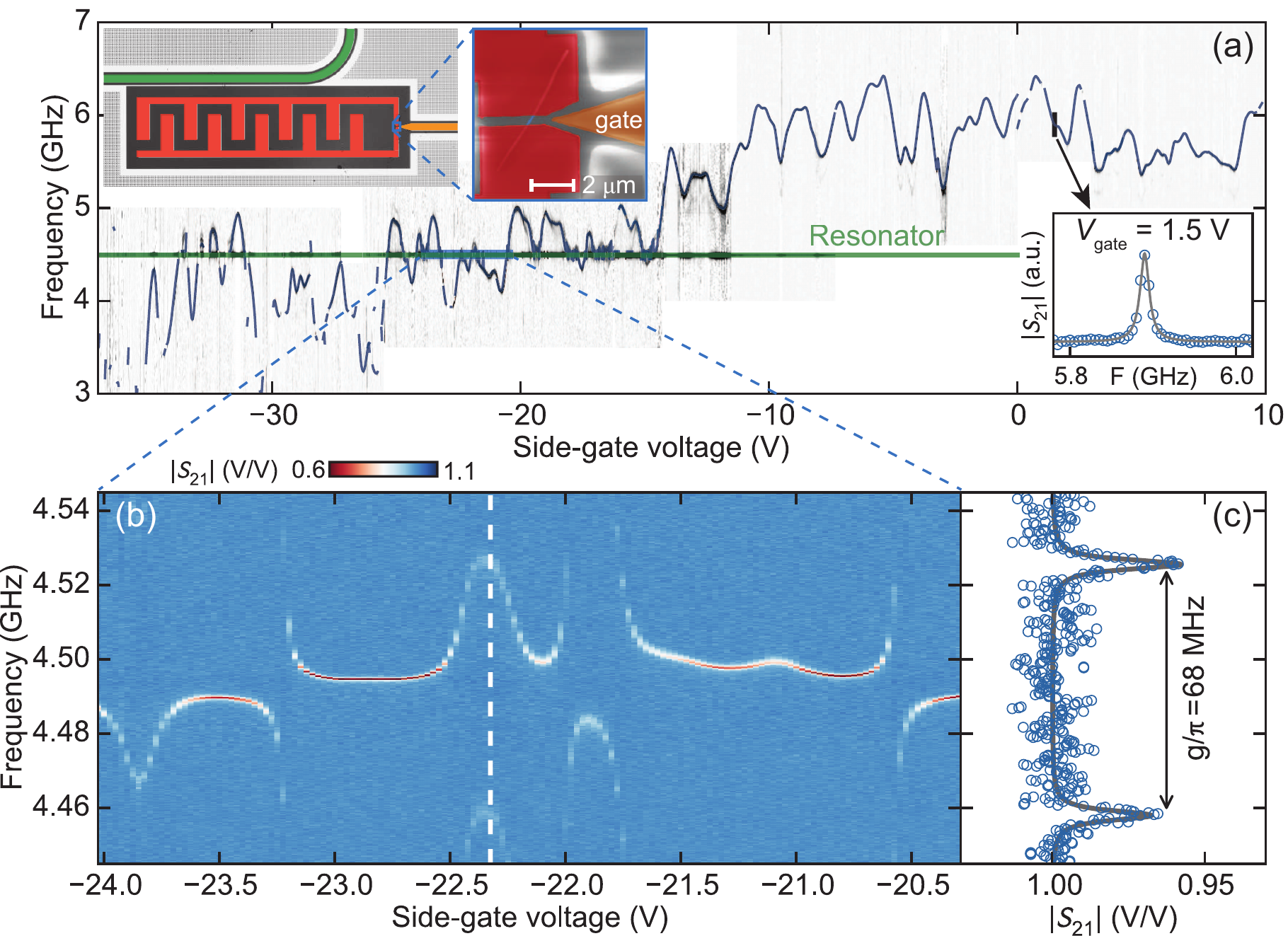}
\caption{Side-gate tuning of a NW circuit and mesoscopic Josephson coupling fluctuations. (a) Left inset; false-colored optical image of Device 2. Right inset; SEM micrograph showing the single NW junction and the proximal side gate (orange) for voltage control. Sweeping this voltage induces reproducible fluctuations in the qubit transition frequency $\fge$. Lower inset; example spectroscopy of the qubit transition, showing an inhomogeneously broadened linewidth $\gamma/2\pi = 13.2 \pm 0.3~\MHz$.
A downward trend in $\fge$ is observed as $\Vg$ decreases. At $\Vg<-15~\V$, $\fge$ fluctuates around the resonator fundamental (green line). (b) A zoom-in around $\Vg = -22~\V$ shows multiple avoided crossings. (c) At $\Vg = -22.3~\V$, the NW circuit fully hybridizes with the resonator. From the minimum splitting, we extract the NW circuit-resonator coupling strength $g/2\pi = 34~\pm~1~\MHz$. }
\end{figure}

We first investigate the electric-field effect on the NW-circuit spectrum [Fig.~2(a)]. Device 2 has one NW junction (measured length $L\approx  550~\nm$) and a proximal side-gate electrode for tuning the carrier density in the NW.
We observe fully reproducible fluctuations~\cite{SOMPRL} in $\fge$ as a function of the side-gate voltage $\Vg$, indicating phase-coherent diffusive charge transport in the NW~\cite{Lee85,Doh05}.
Using the plasma-oscillation relation~\cite{Koch07}, $\Ej \approx \fge^2/8\Ec$, we determine the root-mean-square Josephson-energy fluctuation $\sqrt{\langle \delta^2 \Ej\rangle}/h \approx 2~\GHz$ in the $\Vg = 0-10~\V$ range.
Matching this scale to the Thouless energy~\cite{Beenakker91}, $\Eth = \hbar D/L^2$, and assuming highly transparent contacts~\cite{Abay14}, we estimate the diffusion constant $D \approx 40~\cm^2/\s$. This value is typical for InAs wires~\cite{Jespersen09}.

Side-gate tuning of the NW junction offers a new means to control the spectrum of transmons. Decreasing $\Vg$ brings $\fge$ into resonance with the resonator, revealing multiple avoided crossings [Fig.~2(b)].
The minimum splitting indicates $g/2\pi = 34 \pm 1~\MHz$ [Fig.~2(c)]. We note that while  we only perform quasistatic field-effect tuning of $\fge$ throughout this experiment, nanosecond control should be possible by increasing the bandwidth of off-chip filtering.

We now discuss the impact of charge fluctuations on the observed linewidth of the $\fge$ transition, which is of interest for qubit applications.
Transmon qubits are by design insensitive to charge offset fluctuations on the superconducting islands~\cite{Koch07}, owing to the exponential suppression of charge dispersion when $\Ej\gg\Ec$.
Field-effect control of the Josephson coupling can make $\fge$ sensitive to nearby fluctuating charges. One may expect a region with $\partial \fge/\partial \Vg=0$ to constitute a charge sweet-spot~\cite{Vion02} and thus to correlate with a linewidth reduction. However, we do not observe a correlation between the linewidth $\gamma/2\pi > 10 ~\MHz$ and $\left | \partial \fge/\partial \Vg \right|$, suggesting a different dominant decoherence channel~\cite{SOMPRL}. We surmise a connection between this broad linewidth and the soft gap induced in NWs~\cite{Takei13} contacted by NbTiN using similar fabrication techniques~\cite{Mourik12}. Parallel experiments by the Copenhagen group achieve hard induced gaps in epitaxial Al-InAs NWs~\cite{Chang14} and  $\sim 1~\us$ coherence times in a single-JE hybrid transmon~\cite{Larsen14}. Establishing a clear connection between soft induced gap and dissipation in microwave circuits will be the focus of future studies.   

\begin{figure}[h]
\center
\includegraphics[width=\columnwidth]{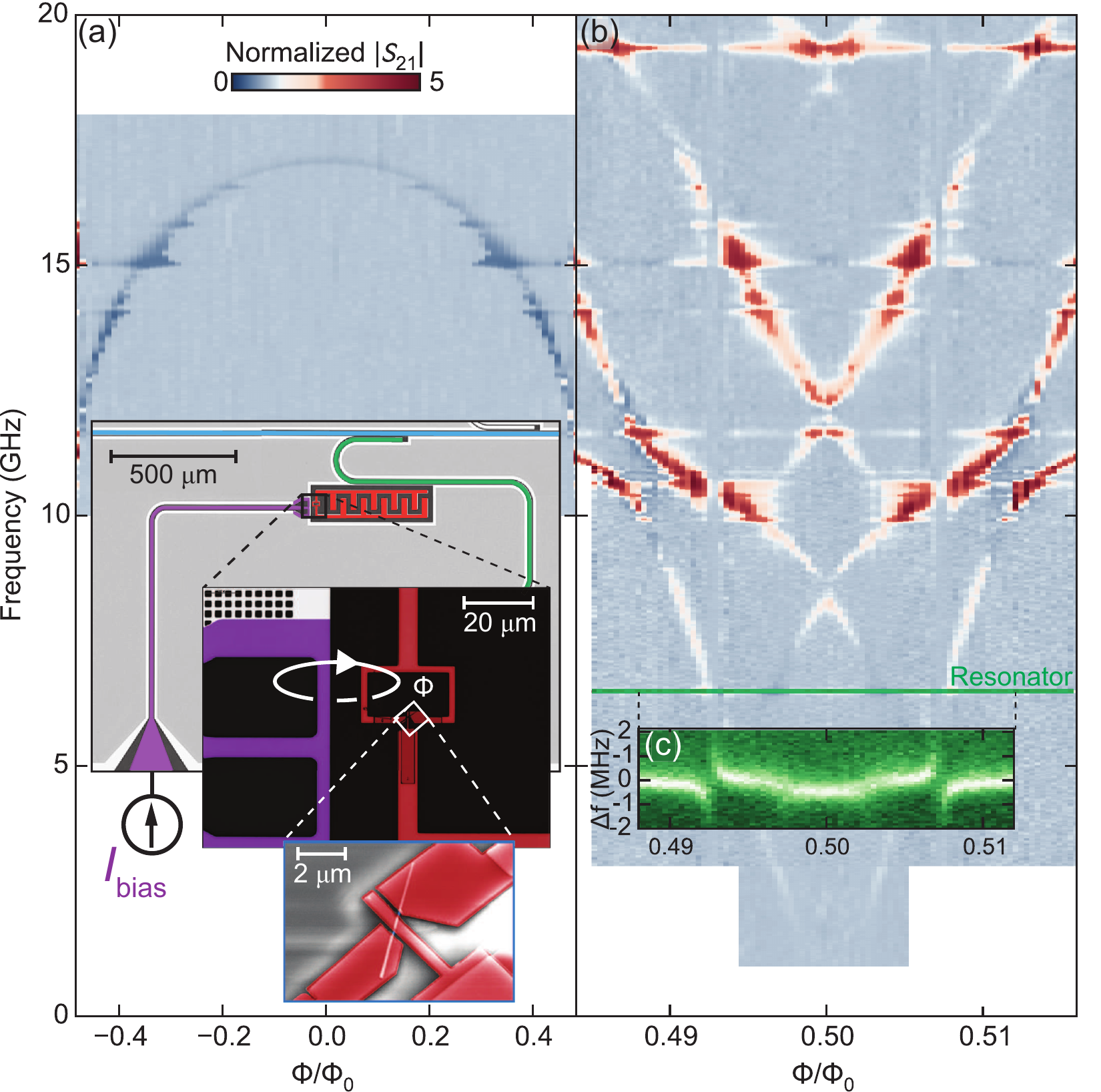}
\caption{Flux-bias spectroscopy of a split-junction NW circuit. (a) Inset: false-colored optical image showing Device 3 (red), its resonator (green), and flux-bias line (purple). Bottom inset: SEM micrograph of the two JEs made from one NW. Flux-bias spectroscopy shows the tuning of $\fge$ with $\Phi$. (b) A high-resolution sweep around $\Phi=\Phi_0/2$ shows a strong flux dependence of the NW circuit transitions. (c) Measurement of resonator transmission around $\fc$ with same horizontal range as in (b). The avoided crossing of the lowest transition with the resonator reveals a much reduced coupling strength.}
\end{figure}

Next, we consider a split-junction device where the two parallel JEs (each with $L=150~\nm$) are created from one $5~\um$ long NW [Fig.~3(a)]. As in conventional transmons, $\fge$ first decreases as flux $\Phi$ is threaded through the loop. However, near $\Phi\sim \Phi_0/2$ ($\Phi_0=h/2e$ is the flux quantum), a clear departure from transmon-like behavior is observed [Fig.~3(b)]. Multiple strongly flux-dependent transitions and a new, strong avoided crossing appear symmetrically about $\Phi_0/2$. In addition, the avoided crossing between the lowest transition and the resonator is strongly reduced [Fig.~3(c)] compared to that of Device 2~\cite{Footnote1DMON}. The highly anharmonic spectrum and much reduced coupling to the resonator shows that the states associated with the lowest-energy transition are no longer plasma modes. 

\begin{figure}[h]
\center
\includegraphics[width=\columnwidth]{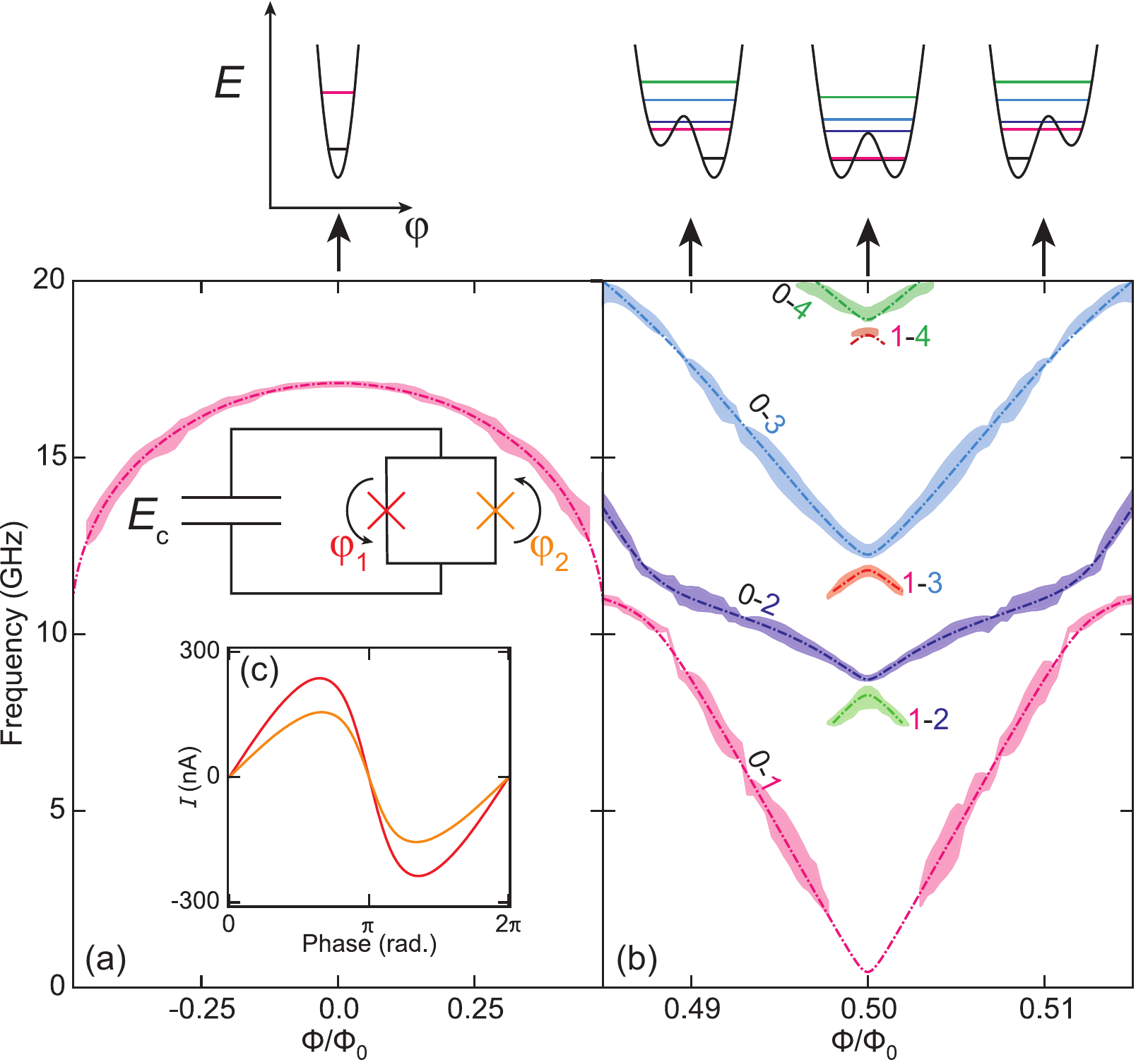}
\caption{Fitting theory to the spectrum of the split-junction NW circuit. (a) Top inset; equivalent circuit of the device, a split Cooper-pair box containing a loop interrupted by two NW JEs (phase differences $\varphi_{1,2}$), threaded by an externally applied flux $\Phi$, and shunted by a capacitance, giving total charging energy $\Ec$. Around $\Phi=0$, the Josephson potential has a single minimum, producing a weakly anharmonic spectrum. (b) The non-sinusoidal current-phase relation of the NW JEs determines the particular flux-dependence of the transition frequencies around $\Phi=\Phi_0/2$ as the Josephson potential develops a symmetric double-well profile that tilts as $\Phi$ is tuned away from $\Phi_0/2$. All curves are the result of a least-squares non-linear fit of the theoretical model described in the text. We identify four fundamental transitions from the ground state and three transitions from the first-excited state. (c) The \cpr~calculated from the fit of the spectrum to the theoretical model described in the text.}
\end{figure}

The observed deviation from the conventional transmon energy spectrum provides a signature of non-sinusoidal \cpr s~in the NW junctions~\cite{Golubov04}. We now show that the observed spectrum can be fully explained by the Hamiltonian of a Cooper-pair box (CPB): $H=4 \Ec\hat{N}+ V_1(\hat\delta) + V_2(2\pi \Phi/\Phi_0-\hat\delta)$,  provided its split junctions do not follow a cosine-shaped Josephson potential [Fig.~4(a)]. Here, the operators $\hat{N}$ and $\hat\delta$ represent the charge imbalance between islands and the phase difference across NW junction 1, respectively. The Josephson potential $V_{i}(\varphi_{i})$ of junction $i$  is linked to its \cpr~by $I_{i}(\varphi_{i})=\left(2\pi/\Phi_0\right)\partial V_{i}/\partial \varphi_{i}$, where $\varphi_1\equiv\hat\delta$ and $\varphi_2\equiv2\pi \Phi/\Phi_0-\hat{\delta}$. Crucially, we require $V_{i}$ to be $2\pi$-periodic but not necessarily cosine shaped. Using a simple phenomenological model~\cite{Beenakker91, SOMPRL} of the form $V_{i}(\varphi_{i})= -K_i\sqrt{1-T_{i}\sin^2(\varphi_{i}/2)}$ and performing a non-linear least-squares fit with five free parameters, we obtain a quantitative match to all spectral data (best-fit values are $\Ec/h= 279\pm1~\MHz$, $K_1/h= 376 \pm 13~\GHz$, $K_2/h= 233\pm2~\GHz$, $T_1=0.86 \pm 0.02$, and $T_2=0.885\pm0.004$). This simple model stricktly applies only to short weak links having a single conductance channel, we therefore use the fit only to extract information about the \cpr s of the junctions. As shown in Fig.~4(a), the corresponding \cpr s are evidently skewed. A three-parameter fit using $V_{i}$ corresponding to the \cpr~of a short, diffusive point contact in the many-channel limit \cite{Kulik75,Golubov04} showed only slightly worse agreement, as did a truncated Fourier series expansion of $V_{i}$. All approaches produce similar skewed \cpr s~\cite{SOMPRL}. 

Interestingly, this device can be operated in two distinct regimes by tuning $\Phi$. Near $\Phi=0$, it operates like a transmon, whose eigenstates are plasma modes with a weakly anharmonic spectrum. Around $\Phi\simeq\Phi_0/2$, it operates like a flux qubit~\cite{Mooij99} whose two lowest energy levels carry opposite persistent currents $\Ipi{i} = \partial{E_i}/\partial{\Phi}$, which we estimate to be of order $\pm 100~\nA$~\cite{SOMPRL}. The possibility to drive transitions between these distinct persistent-current states using coherent microwaves constitutes a manifestation of macroscopic quantum coherence~\cite{Leggett80} in our NW circuits.

In conclusion, we have realized hybrid microwave circuits made from super-semi NW JEs and characterized them using spectroscopy.
NW circuits offer several advantages over traditional aluminum circuits. First, tuning qubit transitions using the field effect in single NW JE devices offers an attractive alternative to flux biasing of split-junction Al devices. Second, these NW circuits are made exclusively from magnetic-field compatible materials.  Magnetic-field compatible super-semi NW microwave circuits have the potential to open new avenues of research. In particular, very pure solid-state electron spin ensembles (e.g., nitrogen impurities in diamond or phosphorous donors in silicon) could be field-polarized to make coherent quantum memories~\cite{Imamoglu09, Ranjan13} for hybrid quantum processors.
In addition, the microwave circuits realized here may be useful to control and readout Majorana bound states~\cite{Alicea11,Hyart13} in proposed demonstrations of non-Abelian exchange statistics~\cite{Moore91,Read00,Ivanov01,Nayak08}. Immediate next experiments will therefore focus on the study of these circuits in up to $\sim 0.5~\Tesla$ in-plane magnetic fields.

\subsection{Acknowledgments}
\begin{acknowledgments}

We thank L.P.~Kouwenhoven, T.M.~Klapwijk, M.H.~Devoret, C.W.J. Beenakker, and J.E.~Mooij for
  useful discussions and comments on the manuscript. We acknowledge funding by
  Microsoft Corporation Station Q, the Dutch Organization for Fundamental
  Research on Matter (FOM), the Netherlands Organization for Scientific
  Research (NWO), and an ERC Synergy Grant. 
\end{acknowledgments}


\begin{thebibliography}{10}
\bibitem{Footnote1DMON}
In addition, several flux-independent lines appear. We can attribute many of
  these to the fundamentals and higher harmonics of other resonators on the
  chip (they are also observed when studying other devices on this chip) and
  are not considered henceforth.
\end{thebibliography}
\end{document}


\title{Supplementary Materials for:\\
{\bf Realization of microwave quantum circuits using hybrid superconducting-semiconducting nanowire Josephson elements}}

\author{G.~de Lange}
\affiliation{QuTech and Kavli Institute of Nanoscience, Delft University of Technology, 2600 GA Delft, The Netherlands.}

\author{B. van Heck}
\affiliation{Instituut-Lorentz, Leiden University, 2300 RA Leiden, The Netherlands.}

\author{A. Bruno}

\author{D.~J. van Woerkom}

\author{A. Geresdi}
\affiliation{QuTech and Kavli Institute of Nanoscience, Delft University of Technology, 2600 GA Delft, The Netherlands.}

\author{S.~R.~Plissard}
\affiliation{Department of Applied Physics, Eindhoven University of Technology, 5600 MB Eindhoven, The Netherlands.}

\author{E.~P.~A.~M.~Bakkers}

\affiliation{QuTech and Kavli Institute of Nanoscience, Delft University of Technology, 2600 GA Delft, The Netherlands.}
\affiliation{Department of Applied Physics, Eindhoven University of Technology, 5600 MB Eindhoven, The Netherlands.}

\author{A.~R.~Akhmerov}

\author{L.~DiCarlo}
\affiliation{QuTech and Kavli Institute of Nanoscience, Delft University of Technology, 2600 GA Delft, The Netherlands.}

\date{\today}
\pacs{74.45.+c, 62.23.Hj, 84.40.Dc, 85.25.Hv}

\maketitle

\section{Materials and Methods}

The chip was fabricated on a sapphire substrate (single-side polished, C-plane
cut, $430~\um$ thickness). After cleaning the substrate in buffered HF, a NbTiN
film ($80~\nm$ thickness) was sputtered. Ground planes, resonators and IDCs were defined by negative tone electron-beam
lithography and reactive ion etching.

In the next step, InAs NWs of typical length $5-10~\um$ and radius $50-100~\nm$ were
controllably deposited using a micromanipulator setup equipped with an
optical microscope. Detailed structural and DC transport characterization of the
InAs NWs used in this experiment have been published in earlier reports~\cite{Scheffler08}.

\begin{figure*}[!h]

\includegraphics[width=0.7\textwidth]{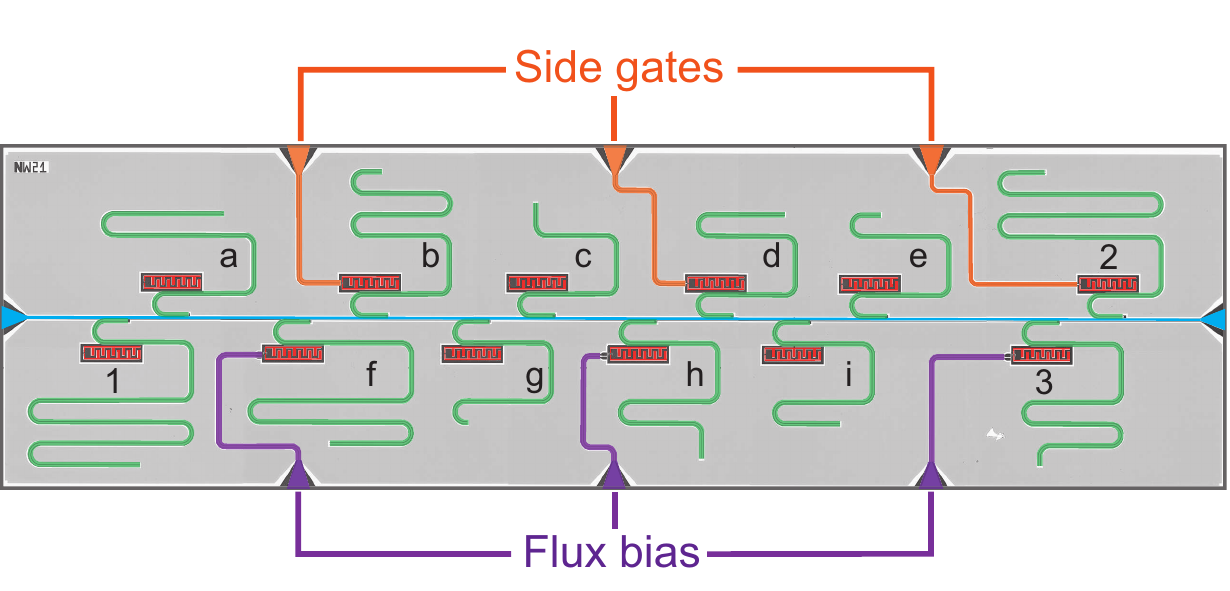}
\caption{{Overview of the chip.}
}
\end{figure*}

The fine patterns were designed based on optical alignment to the NWs. The semiconductor-superconductor interface area was maximized by
covering the NW as much as possible. The overlap between the coarse and
the fine superconducting structures in each IDC was designed to be several tens of $\um^2$.
After defining the fine pattern by positive tone
electron-beam lithography, the semiconductor surface was cleaned in buffered
HF for $20~\s$ and then a $100~\nm$ thick NbTiN film was sputtered
in order to overlay the coarse structures as well.
\begin{figure}
\begin{center}
Table S1. \\Overview of devices on the chip. $R_\mathrm{RT}$ is the two-terminal resistance probed close to the junction at room temperature.\\
  \begin{tabular}{| l | l | l | l |l|l| }

    \hline
    Device & $ R_{\mathrm{RT}}~(\kOhm)$ & $L_\mathrm{design}~(\nm)$ & $f_\mathrm{c}~(\GHz)$& $\fge~(\GHz)$ & Description \\ \hline \hline
    1 & 5.7 & 500& 4 &6.7 & single (Fig.~1)\\ \hline
    2 & 6.2 & 500& 4.5 &6.2 ($\Vg=0~\V$) & single, with gate (Fig.~2)\\ \hline
    3 & 2.2 & 200& 6.5 &17 (max)& split, with flux control  (Figs.~3\&4)\\ \hline
    a & 3.03 & 200& 7 &not $<20$ &single\\ \hline
    b & 3.07 & 200& 6 &13.6& single, with gate\\ \hline
    c & short & 50& 11 &-& single\\ \hline
    d & open & 100& 8.5 &-& single, with gate\\ \hline
    e & short & 50& 10 &-& single\\ \hline
    f & 3.5 & 500& 5 &9.1 (max)& split, with flux control\\ \hline
    g & short & 50& 10.5 &-& single\\ \hline
    h & 1.4 & 100& 8 &not $<20$& split, with flux control\\ \hline
    i & 2.6  & 100& 9 &not $<20$& single\\ \hline
    \hline
  \end{tabular}
\end{center}
\end{figure}
In total, twelve devices were fabricated on the chip (Fig.~S1), and each was coupled to a coplanar waveguide quarter-wave resonator with a distinct fundamental frequency in the range $4$ to $11\,$GHz.
Table~1 gives an overview of all the devices on the chip.
Gate voltages in the range $\pm 40~\V$ were applied through on-chip $50~\Ohm$ transmission lines with open ends proximal to the NW. Off-chip, the lines are filtered by a
second-order RC filter and a copper-powder (CuP) filter mounted to the mixing chamber plate of
the refrigerator (Fig.~S2). The two filters have $\sim 50~\kHz$ and $\sim 1~\GHz$ cutoff frequencies, respectively.  
The JEs of split-junction devices were embedded in a superconducting loop of area $13 \times 24~\um^2$. Flux through the loop is controlled using
on-chip $50~\Ohm$ transmission lines, with short-circuited termination proximal to the loop. Off-chip, the lines are low-pass filtered with nominal $1~\GHz$ cutoff frequency.

\begin{figure*}[!h]

\includegraphics[width=\textwidth]{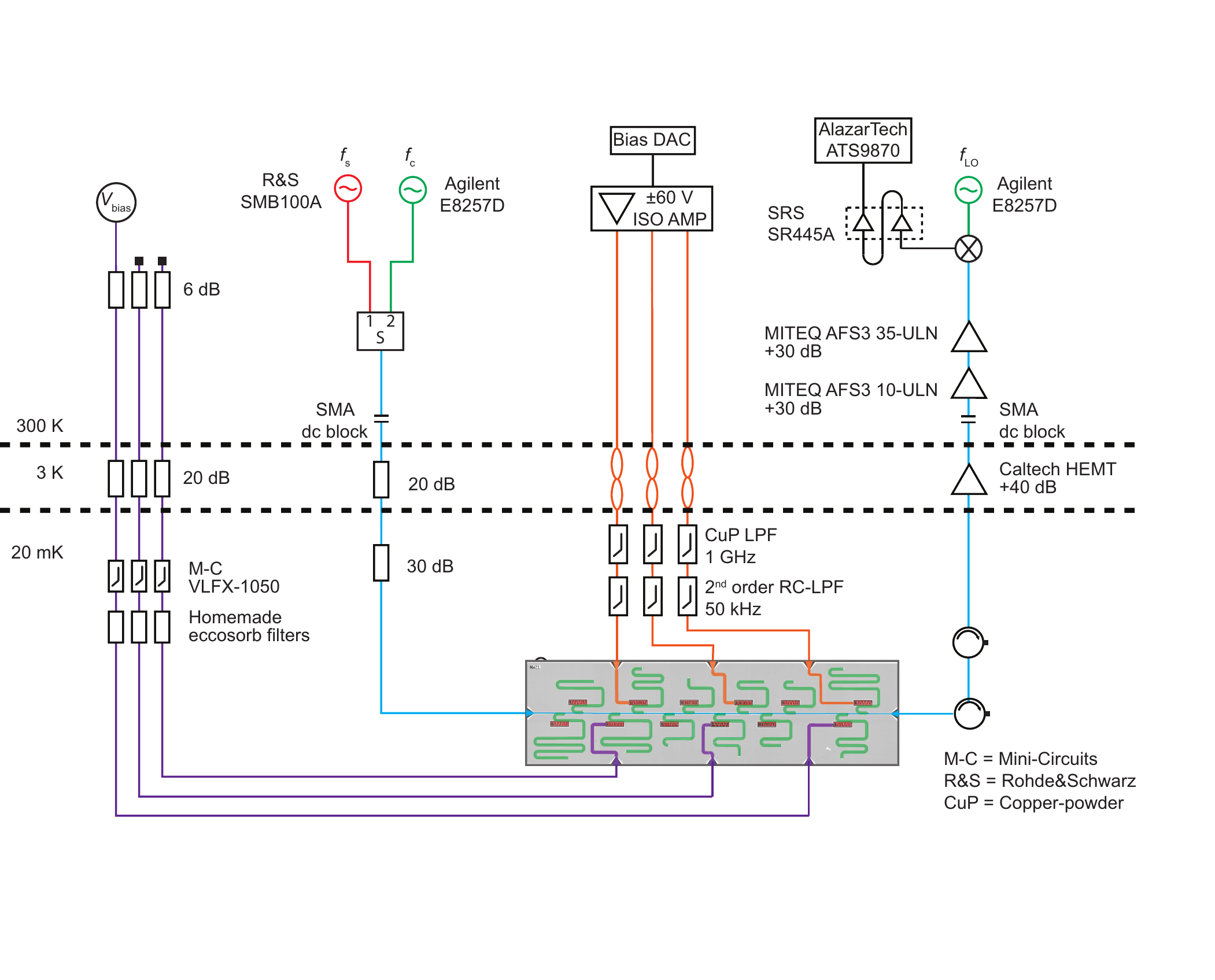}
\caption{ Wiring schematic of the experimental setup outside and inside the
$\Hethree$-$\Hefour$ dilution refrigerator (Leiden Cryogenics CF-650). Readout (green) and spectroscopic (red) microwave drives are combined in a single coax line (blue) at room temperature. The line is attenuated at the 3 K and 20 mK stages before connecting to the chip feedline. The feedline output is isolated from the higher temperature stages by two circulators. The signal is amplified at the 3 K stage by a HEMT amplifier (Caltech Cryo1-12, 0.06 dB noise figure) and at room temperature by two Miteq amplifiers. The signal is downconverted to $1~\MHz$, further amplified, digitized and saved for processing. The setup contains wiring for gate (orange) and flux (purple) control on a subset of devices. Both types of control line are low-pass filtered.
}
\end{figure*}

\clearpage
\section{Additional data for Device~2}

\begin{figure*}[!h]
\includegraphics[width=0.6\textwidth]{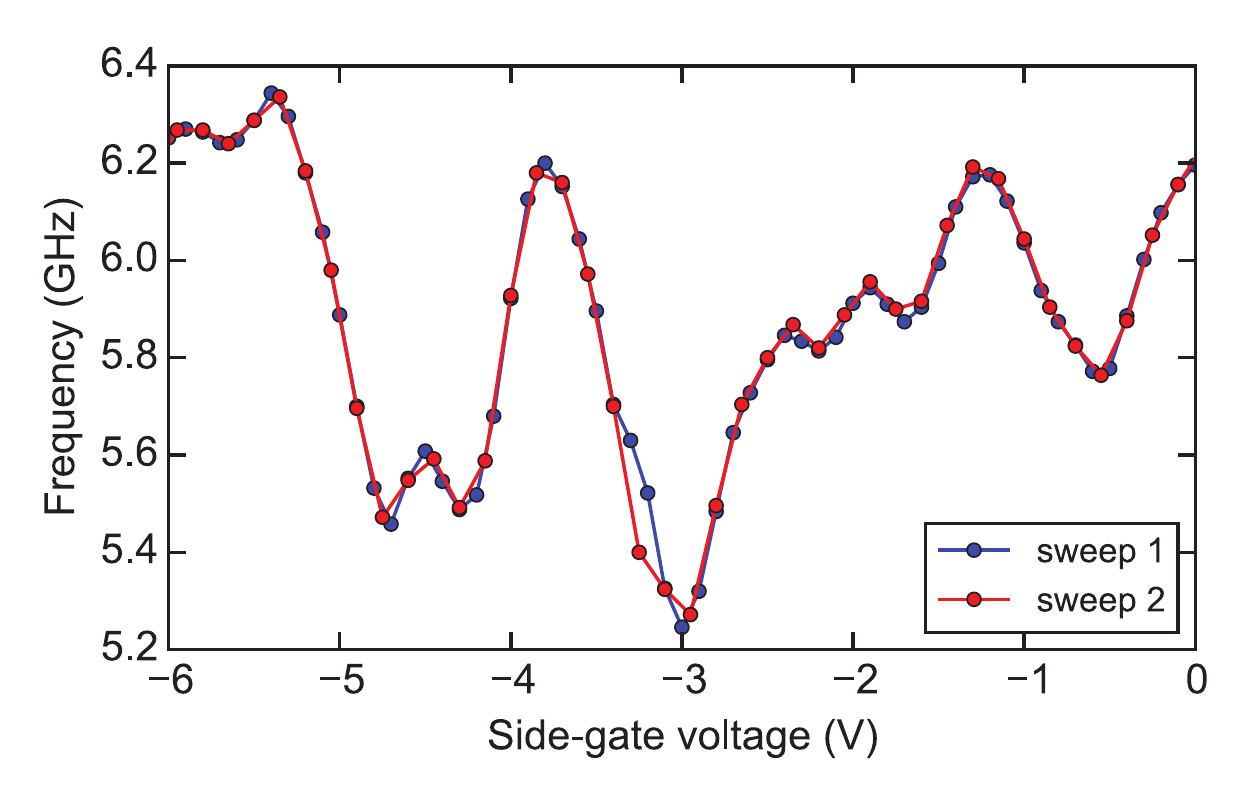}
\caption{{ Reproducibility of the $\fge$ fluctuations}. Two sweeps of two-tone spectroscopy show the reproducibility of the $\fge$ fluctuations. These sweeps were separated by two days.
}
\end{figure*}

\begin{figure*}[!h]

\includegraphics[width=0.9\textwidth]{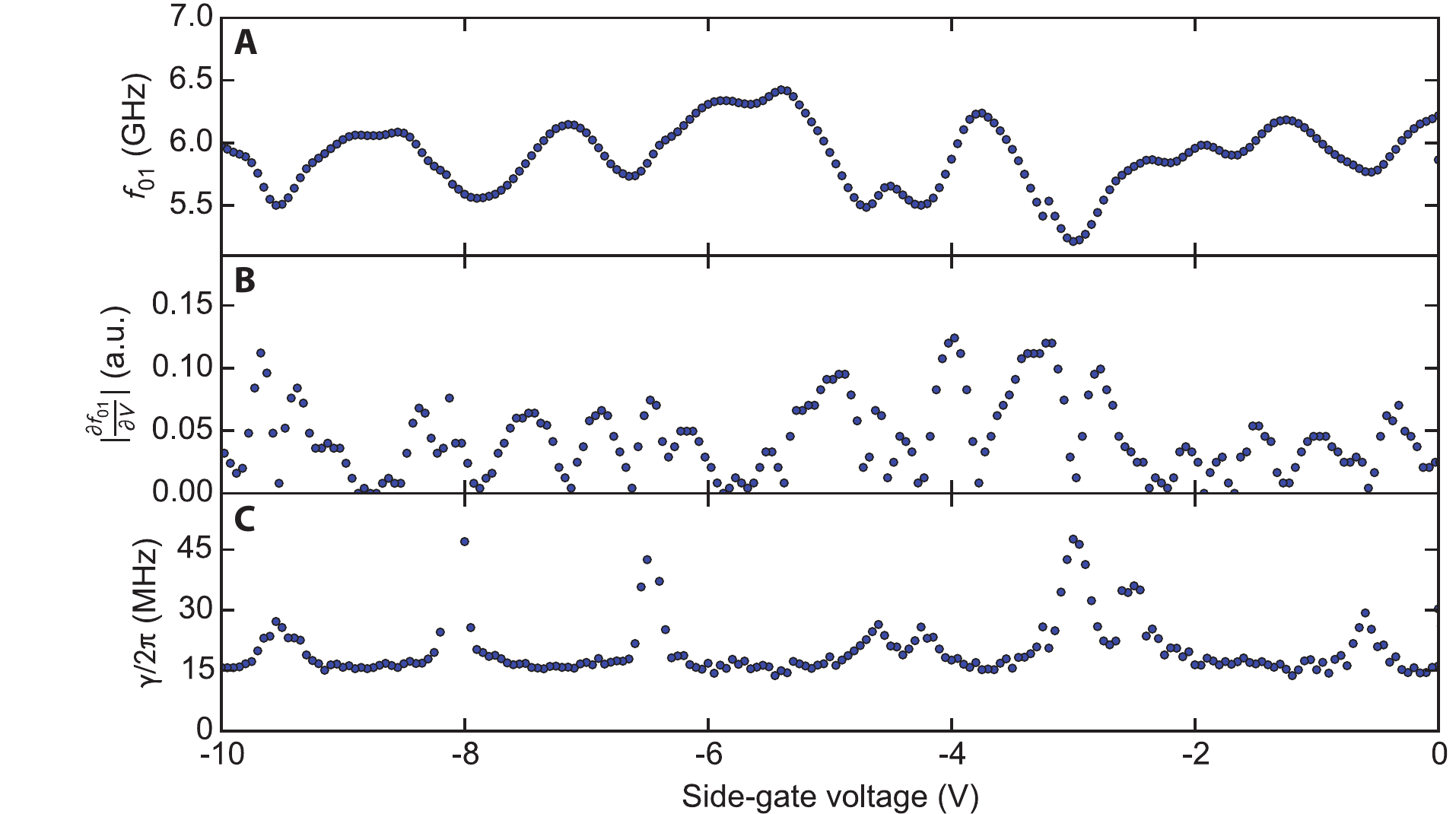}
\caption{{ Absence of correlation between $\fge$ field-effect sensitivity and linewidth $\gamma$}. {\bf (A)} Qubit transition frequency $\fge$ as a function of $\Vg$, extracted from Fig.~2. {\bf (B)} Computed field-effect sensitivity $\left |\partial{\fge}/\partial{\Vg} \right|$. {\bf (C)} Extracted $\gamma$. Vanishing  $\left |\partial{\fge}/\partial{\Vg} \right|$ does not correlate with a reduction in $\gamma$. 
}
\end{figure*}

\clearpage
\section{Data extraction from flux-bias spectroscopy (Fig.~3)}

In order to perform fits to the observed spectroscopy lines, we have extracted a set of data points with error bars from the raw data, consisting of a 2D scan of feedline transmission measured as a function of frequency $f$ and voltage $V_\mathrm{bias}$ applied to the $50~\Ohm$ bias line. We used the following procedure to extract the points~\cite{Footnote2DMON}.
\begin{enumerate}
	\item We applied a Gaussian filter to suppress noise fluctuations, and subtracted the residual background signal.
	\item We converted $V_\mathrm{bias}$ into flux $\Phi$ through the loop, assuming a linear relation $\Phi= A V_\mathrm{bias} + B$.
	\item We identified isolated features in the raw data by removing all points below a transmission threshold. Every isolated feature consisted of a connected set of data points surrounding peaks of high transmission in the $f$-$\Phi$ plane.
	\item Within every feature, we extract a single data point for each voltage value. The frequency $f$ and uncertainty $\Delta f$ of every point was computed by taking the average frequency of all points in the same feature with the same $V$, weighted by their transmission amplitude.  
	This procedure provided a collection of data points $(\Phi, f, \Delta f)$. The points were manually divided into groups forming continuous $f(\Phi)$ transition lines.
\end{enumerate}

\section{Theoretical model and fits for Device~3}
Many features  of the observed transitions coincide with those expected for a double-well potential. The first is the linear vanishing of $\fge$ at $\Phi = \Phi_0/2$, which is consistent with the small energy difference between the ground and first-excited states in a symmetric double-well potential.
The second is the appearance of a strong avoided crossing between the two lowest transitions at $\Phi \simeq 0.49 \Phi_0$.
The avoided crossing naturally arises in a tilted double-well potential, when the lowest energy state in the shallower well becomes resonant with the first-excited state of the deeper well.
The third is the visibility of transitions whose frequency decreases away from $\Phi=\Phi_0/2$, which is consistent with residual thermal population of the first-excited state close to $\Phi_0/2$.

The transition frequencies obtained from the energy levels of the split-junction CPB Hamiltonian,
\begin{equation}
H=4 \Ec\hat{N}+ V_1(\hat\delta) + V_2(2\pi \Phi/\Phi_0-\hat\delta)\,.
\end{equation}
were fit to the extracted data points. The energy levels were computed numerically in the eigenbasis of the charge operator $\hat{N}$, truncating the Hilbert space to $\sim 200$ states such that convergence was reached. Via a Fourier series expansion of the $2\pi$-periodic functions
\begin{equation}
V_i(\varphi_i) = -\sum_{n} A_{i,n} \,\cos(n \, \varphi_i)\,,
\end{equation}
the Josephson terms in the Hamiltonian can be easily expressed in the same basis using the raising and lowering operators $\hat{N}_{\pm}=\exp \,(\pm i\hat{\delta})$,
\begin{eqnarray*}
V_1(\hat\delta) &= &-\sum_{n} A_{1,n} (\hat{N}_+ + \textrm{h.c.}),\\
V_2(2\pi \Phi/\Phi_0-\hat\delta) &=& -\sum_{n} A_{2,n} (\hat{N}_+e^{-i2\pi \Phi/\Phi_0}  + \textrm{h.c.})\,.
\end{eqnarray*}
Hence, the Fourier series of $V_{i}$ contains higher harmonics with period $2\pi/n$, with $n>1$, allowing the total Josephson energy $V_1+V_2$ to develop several local minima in a $2\pi$ interval, thereby forming a double-well potential.
 This situation is particularly relevant at $\Phi\simeq \Phi_0/2$, where the odd harmonics of $V_1$ and $V_2$ subtract. Time-reversal symmetry
makes the minima degenerate at $\Phi=\Phi_0/2$ (Fig.~4).

Lacking exact knowledge of the \cpr~of the junctions, we have tried different phenomenological models for the $V_i(\varphi_i)$:
\begin{enumerate}
	\item The model presented in the text,
	\begin{equation}
	V_i(\varphi_i)= -K_i\sqrt{1-T_{i}\sin^2(\varphi_{i}/2)},
	\end{equation}
	is strictly valid only for short junctions, with total amplitude $K_i \simeq n_i \Delta_0$, with $\Delta_0$ the induced gap in the NW, resulting from $n_i$ transport channels with a typical transparency $T_{i}$.
	This model leads to a Hamiltonian with five free parameters: $\Ec$, $K_1$, $K_2$, $T_1$, and $T_2$. For the fit we took the Fourier series expansion of $\V_i$ and truncated to seven harmonics in the numerical diagonalization of the Hamiltonian. 
	\item The Kulik-Omelyanchuk (K-O) model for the zero-temperature current-phase relation of a diffusive and short point contact~\cite{Kulik75,Golubov04} gives
	\begin{equation}
	V_i(\varphi_i) = -2 K_i\int_0^{\pi/2}\sqrt{1-\sin^2(\varphi_i/2) \sin^2 x}\, \mathrm{d} x.
	\end{equation}
	The above relation can be obtained from Eq.~(3) by integrating over the probability distribution of transmission coefficients $T$ in a diffusive point contact. This model leads to a Hamiltonian with three free parameters, $\Ec$, $K_1$, and $K_2$. As before, the Fourier series was truncated to seven harmonics. The best-fit results are shown in Fig.~S5.
	\item We have also performed a  direct fit of the Fourier coefficients $A_{i,n}$, increasing the number of coefficients until a good agreement with the data was observed. The fit is more sensitive to the Fourier coefficients of the weak junction, as the asymmetry in junction strength makes the phase difference across the strong junction stay close to zero at any $\Phi$. The  best-fit values of the seven-parameter fit are $\Ec/h= 275\pm 1~\MHz$, $A_{1,1}/h  = 95.5\pm0.4~\GHz $, $A_{2,1}/h = 68.9\pm0.2~\GHz$, $A_{2,2}/h = -11.78\pm0.06~\GHz$, $A_{2,3}/h = 3.28\pm0.02~\GHz$, $A_{2,4}/h =  -0.98\pm0.03~\GHz$, and $A_{2,5}/h =  0.21\pm0.03~\GHz$. Including more Fourier coefficients did not improve the agreement with the data, and increased the variance of the best-fit values.
\end{enumerate}

\begin{figure*}[!h]

\includegraphics[width=0.7\textwidth]{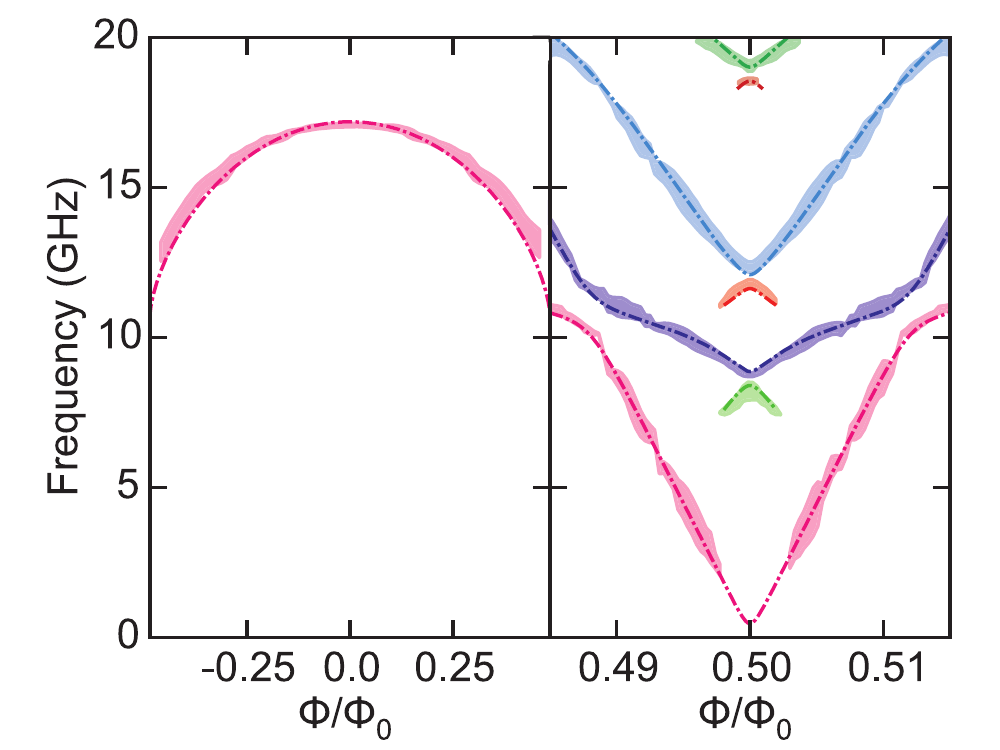}
\caption{{ Fit to the K-O model.} Best-fit values are $\Ec/h = 266 \pm 2~\MHz$, $K_1/h = 218\pm1~\GHz$, and $K_2/h = 141\pm1~\GHz$.
}
\end{figure*}

\begin{figure*}[!h]

\includegraphics[width=0.5\textwidth]{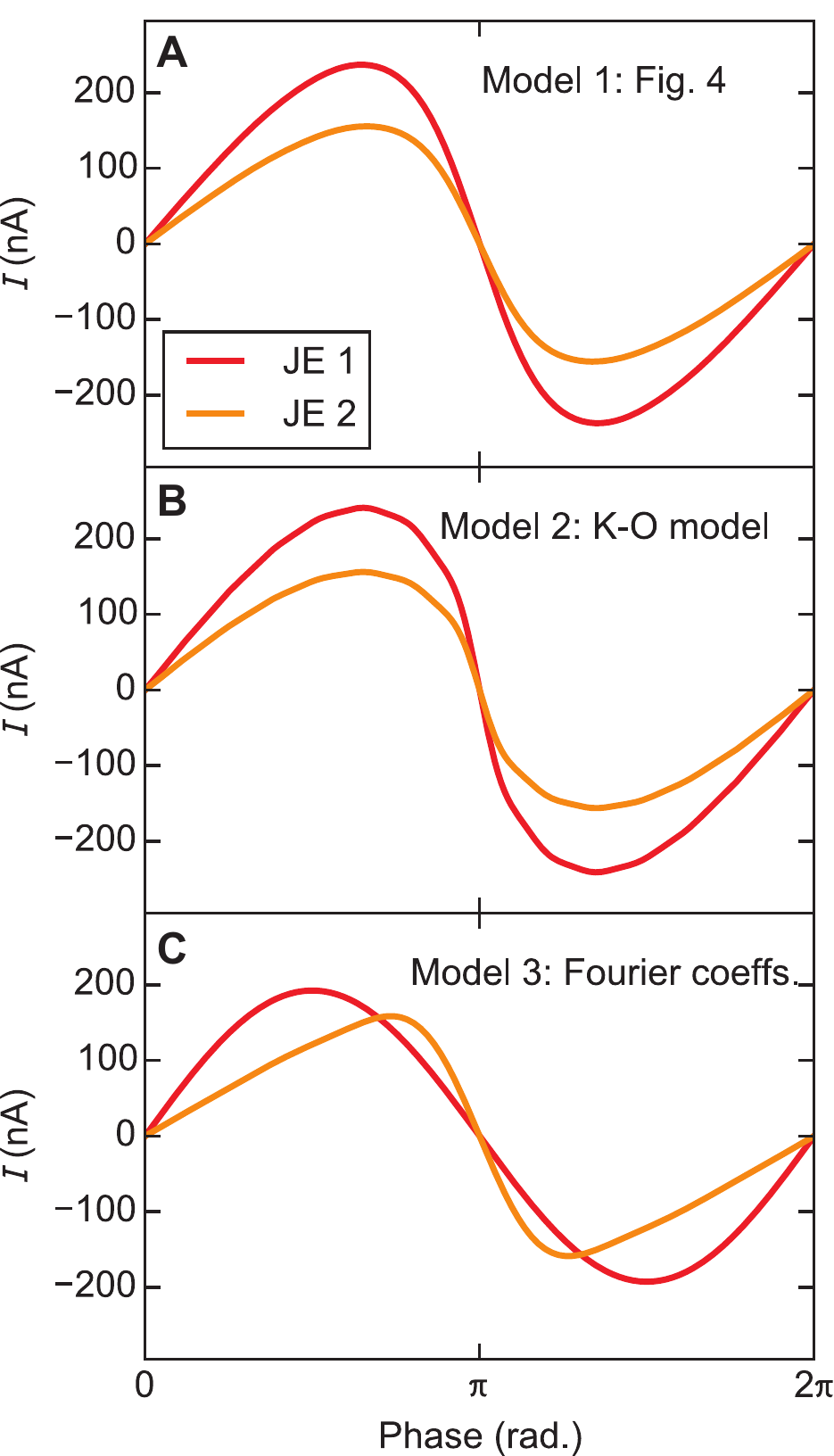}
\caption{{ \cpr s for the three models.} {\bf (A)} Phenomenological model used in the main text. {\bf (B)} K-O model , {\bf (C)} Fourier series using one harmonic for JE 1 and five harmonics for JE 2.
}
\end{figure*}

Due to the non-linearity of the fitting model and correlated errors in the extracted data points, it is not straightforward to extract a figure of merit for quantitative comparison of the models. All models require at least one non-sinusoidal  \cpr~to fully explain all the features of the data set.

Although the data are well fit by non-sinusoidal \cpr s, there are other mechanisms that can produce similar spectra in superconducting circuits. First, the presence of a superinductance is known to produce a potential landscape with several minima, as for the fluxonium~\cite{Manucharyan09}.
NbTiN is known for its high kinetic inductance $\Lk$. Indeed, in our circuit we estimate $\Lk \approx 0.2~\nH$ and loop geometric inductance $\Lg \approx 30~\pH$.
The associated total inductive energy $\El = (\Phi_0/2\pi)^2/\Lk \approx h \times 800~\GHz$ is, however, much larger than $\Ej \approx (K_1 T_1+K_2 T_2)/4 = h \times 132~\GHz$ and therefore cannot produce multiple minima (fluxonium requires $\El<\Ej$).
Second, microwave-induced excitation of Andreev bound states (ABS) confined to the junction can also produce a qualitatively similar spectrum~\cite{Bretheau13}. 
However, ABS transitions below $2~\GHz$  require almost perfectly transparent transport channels, which has proven challenging in even the most ideal atomic point contacts~\cite{Bretheau13}.
We therefore surmise that the JEs remain in their ABS ground state and that the observed spectrum is fully due to plasma modes. The potential to observe ABS transitions provides an exciting prospect for future work.

\section{Estimation of Device~3 parameters from model}

It is possible to estimate interesting properties of Device~3 using the best-fit model values presented in the main text. In particular, we can compute the dipolar couplings of the NW circuit to its resonator.
The coupling strength of the dipole-induced transition between two eigenstates $\ket{k}$ and $\ket{l}$ of the split-junction CPB Hamiltonian is proportional to the corresponding matrix element of $\hat{N}$, $n_{kl} = \bra{k}\hat{N}\ket{l}$.

Fig.~S7 shows the calculated relative couplings $ \left| n_{01}\right|$, $ \left| n_{02}\right|$, $ \left| n_{03}\right|$, and $ \left| n_{12}\right|$ near $\Phi=\Phi_0/2$. Their flux dependence explains some interesting features of the data, in particular:
\begin{enumerate}
	\item The vanishing visibility of the $0\to2$ transition at $\Phi=\Phi_0/2$.
	\item The strong visibility of the $0\to3$ transition near $\Phi = \Phi_0/2$.
	\item The reduction of $\left| n_{01}\right|$ near $\Phi_0/2$ (see Fig.~3) compared to that of Device 2.
\end{enumerate}

\begin{figure*}[!h]

\includegraphics[width=0.7\textwidth]{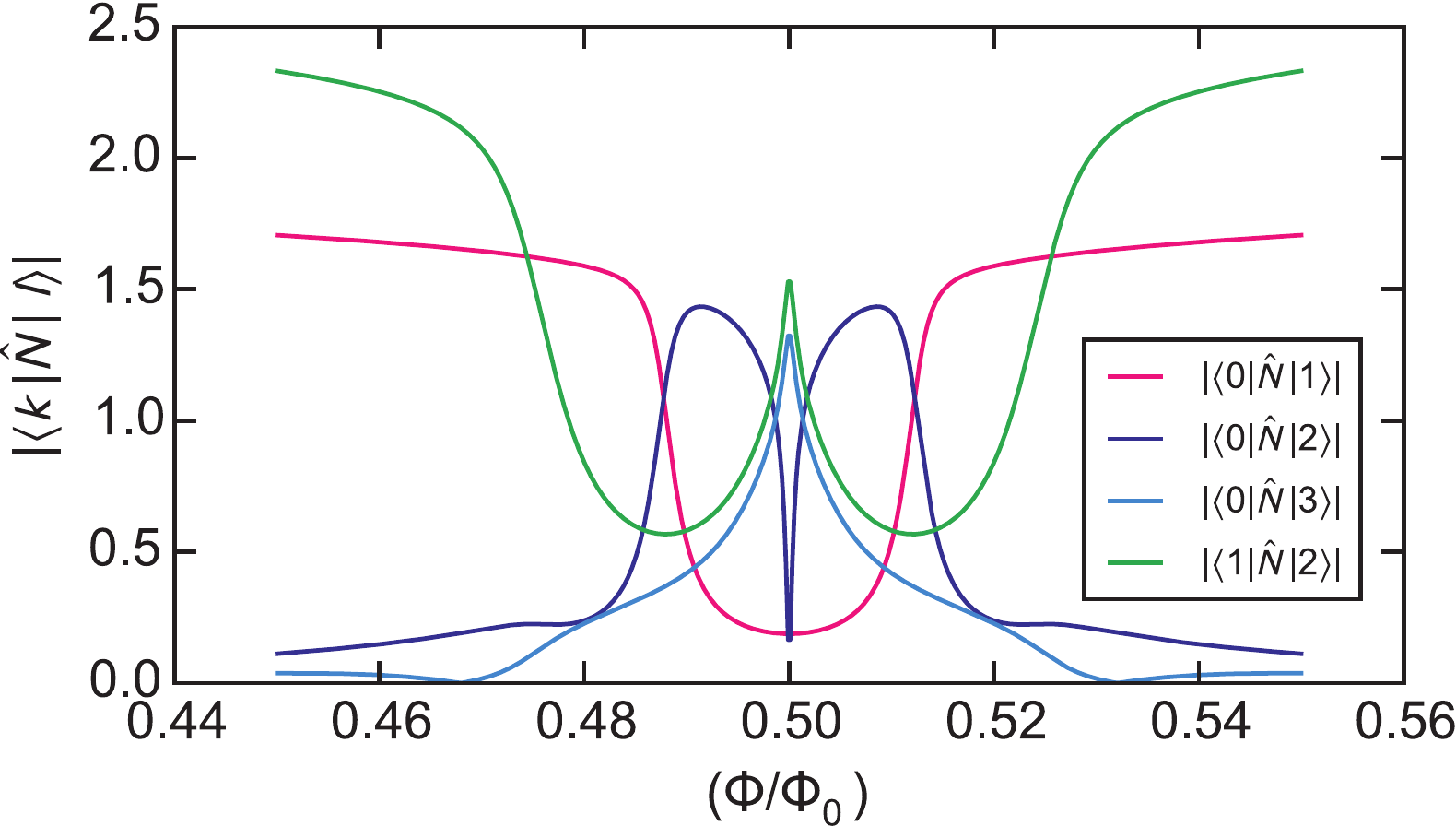}
\caption{{ Flux-bias dependence of the NW-circuit-resonator coupling matrix elements for Device~3.} The absolute matrix element $\left|\bra{k}\hat{N}\ket{l}\right|$ determines the visibility of the transitions in Fig.~3 of the main text. As expected for transmons, only the matrix elements $\left|\bra{k}\hat{N}\ket{k+1}\right|$ remain strong away from $\Phi = \Phi_0/2$.
}
\end{figure*}

\begin{figure*}[!h]

\includegraphics[width=0.7\textwidth]{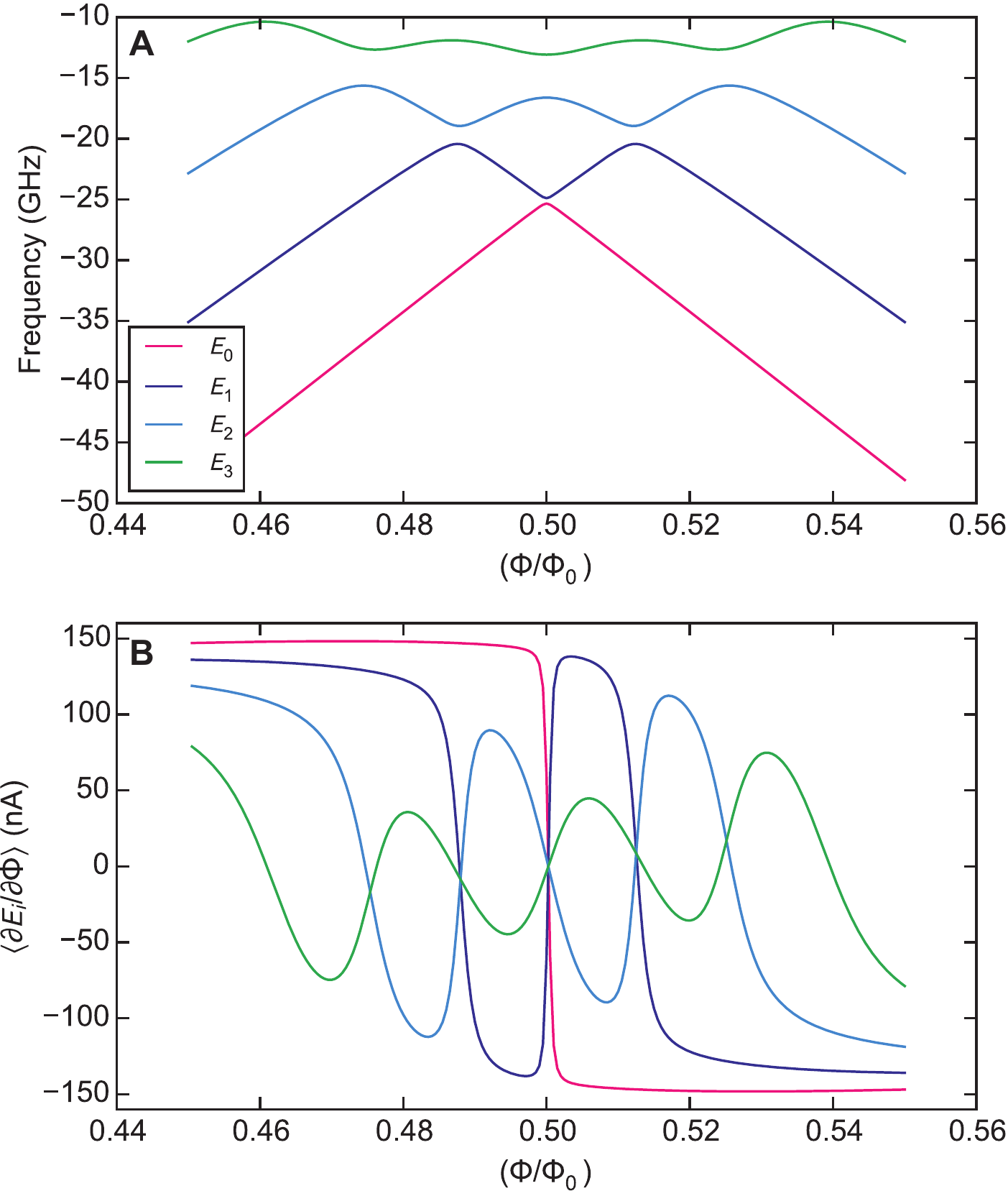}
\caption{{ Energy levels of the split-junction NW circuit around $\Phi_0/2$.}  {\bf (A)} Calculated energy levels for the parameters of Fig.~4 of the main text. {\bf (B)} The states belonging to the lowest energy levels, $E_0$ and $E_1$, carry opposite persistent currents $I_{\mathrm{p},i} = \partial{E_i}/\partial{\Phi}$. The two states hybridize at $\Phi=\Phi_0/2$. Driving $\fge$ with microwaves induces transitions between these two macroscopically distinct current states.
}
\end{figure*}